\documentclass[aps,prb,preprint,superscriptaddress,showpacs,floatfix]{revtex4}
\usepackage{graphicx}
\usepackage{amsmath}

\begin{document}
\title{Perturbative calculation of non-local
corrections to dynamical mean field theory}
\author{V. I. Tokar}
\affiliation{Institute of Magnetism, National Academy of Sciences,
36-b Vernadsky street, 03142 Kiev-142, Ukraine}
\author{R. Monnier}
\affiliation{Laboratorium f{\"u}r Festk{\"o}rperphysik, 
Eidgen{\"o}ssische
Technische Hochschule-H{\"o}nggerberg, 8093 Z{\"u}rich, Switzerland}
\date{\today}
\begin{abstract}
A technique allowing for a perturbative treatment of nonlocal
corrections to the single-site dynamical mean-field theory (DMFT) in
finite dimensions is developed.  It is based on the observation that
in the case of strong electron correlation the one-electron Green's
function is strongly spatially damped so that its intersite matrix
elements may be considered as small perturbations.  Because the non-local
corrections are at least quadratic in these matrix elements, DMFT in
such cases may be a very accurate approximation in dimensions $d=$ 1--3.
This observation provides a rigorous justification for the application
of DMFT to physical systems.  Furthermore, the technique allows for a
systematic evaluation of the nonlocal corrections. This is illustrated
with two explicit examples. First we calculate the magnetic short
range order parameter for nearest neighbor spins in the insulating phase
of the half filled Hubbard model on the square lattice which exhibits
an excellent agreement with the results of a recent cluster approach.
As a second example we study the lowest order correction to the
DMFT self-energy and its influence on the local density of states.
\end{abstract}
\maketitle
\section{\em Introduction}
The description of systems of strongly correlated electrons
provided by the dynamical mean field theory (DMFT) has
significantly improved our understanding of this class of
materials.\cite{Metzner_Vollhardt89,Georges_ea96,Held_ea97,Imada_ea98}
The major limitation of DMFT lies in  its single-site
nature,~\cite{Georges_ea96,Held_ea97} which makes it unable to account
for spatial correlations in finite-dimensional systems. To remedy
this deficiency, cluster generalizations of DMFT have  been devised
(see discussions on this subject in Refs.\ \onlinecite{Maier_ea05b}
and \onlinecite{Okamoto_ea03}).  However, in order to recover the
thermodynamic limit, clusters of sufficiently large size should be
used.\cite{Maier_ea05b} This strongly enhances the numerical effort needed
to solve the corresponding DMFT equations, because the number of quantum
variables to be simulated grows proportionally to the number of atoms
in the cluster.  To alleviate this difficulty, less costly but less accurate 
numerical approaches have been
proposed.\cite{Garcia_ea04,Dai_ea05,Okamoto_ea05} In our
opinion, this is not a viable solution, because the reduction of errors brought
about by the larger number of atoms in the cluster can be overcompensated
by the cruder treatment of the quantum effects.

The aim of the present paper is to propose a perturbative approach
accounting for nonlocal corrections to  DMFT, with the latter being
considered as the zeroth order approximation.  The approach derives from
the general method proposed in Ref.\ \onlinecite{Tokar85}. This so-called
gamma-expansion method (GEM) proved to be successful, {\em inter alia}, in
the treatment of the local atomic correlations or short range order in the
electronic theory of alloys.  \cite{Masanskii_ea88,Monnier_ea98}. Here we
apply it to the Hubbard model on a square lattice and in the insulating
phase, which is well described by the alloy analogy (or Hubbard III
approximation) \cite{Hubbard64,Georges_ea96}, and therefore should be
amenable to the same technique.  This will be illustrated below with an
explicit calculation of the magnetic short range order (SRO) for this
model.  The non-local magnetic correlations are of paramount importance
for the theories of high temperature superconductivity based on the spin
fluctuation mechanism,\cite{Scalapino95}  and their enhancement should
therefore be of interest in these theories.  In particular, both the Neel
and the superconducting transition temperatures should monotonically
depend on the strength of the short range magnetic correlations.
In this respect the observation made in Ref. \onlinecite{Okamoto_ea05}
that different cluster theories give short range correlations differing
roughly by a factor of two provides  an additional reason for developing
non-cluster based treatments of non-local correlations.

The proposed expansion is based in
part on the widely accepted opinion that the single-site DMFT which is
exact in infinite dimensions provides us with a picture of the strong
correlation phenomena which is qualitatively valid also in physical
dimensions $d=$1--3.  In other words, it is the single-site dynamics
which contain the essence of the problem while intersite correlations
play lesser role and so may be considered as correction terms.

An approximate non-local theory in finite dimensions, in which
the single-site DMFT is used as the source of information on
the local correlations has recently been proposed
by Kusunose\cite{Kusunose06} and Toschi et al.\cite{Toschi_ea07}.
In essence, the method consists in first finding {\em local} irreducible vertex
functions from the full DMFT correlators and then using these in
the Bethe-Salpeter\cite{Kusunose06} or parquet\cite{Toschi_ea07} equations
coupled with Dyson-like equations for the electron self-energy
with {\em nonlocal} electron propagators in
order to find the {\em nonlocal} self-energy.
From our point of view the main deficiency of such approaches is the
lack of consistency: The electron
propagators in a skeleton expansion are the same in
side-pieces and inside the crosspieces of a ladder diagram.  
Taking into account
that for large values of the coupling constant there is no reasons for
the smallness of the terms in the ladder sums, in a strict mathematical sense
the series are divergent.  So the neglect of non-local
contributions in some parts of the diagrams not only is unfounded
but may also lead to serious numerical errors.   In Ref.\
\onlinecite{Kusunose06} a reference to the spirit of the DMFT in
infinite dimensions was
invoked where the irreducible vertices inside the ladder series 
can be chosen to be local without loss of 
accuracy.\cite{Zlatic_ea90,Georges_ea96}  This, however, is completely
due to specifics of the $d=\infty$ case, while in finite dimensions
the leading nonlocal corrections are are proportional to
$O(1/d^{1/2})$\cite{Toschi_ea07} and thus are quite large in physical dimensions $d=$1--3.
In our approach, however, the smallness of the non-local contributions
can be explicitely established  {\em a priori}.  

\section{\em The model}
We are going to study the Hubbard model defined by the Hamiltonian
\begin{equation}
\label{ H}
\hat{H}=-\sum_{i,j, s}t_{ij}\hat{c}^{\dagger}_{i s}\hat{c}_{j s}
+U\sum_i\hat{n}_{i\uparrow}\hat{n}_{i\downarrow}
\equiv H_0+H_U,
\end{equation}
where the number operators $\hat{n}_{i s}$ for electrons with the spin
projection $ s=\uparrow,\downarrow$ are expressed through the creation
and annihilation operators $\hat{c}^{\dagger}_{i s}$ and $\hat{c}_{i
s}$ as: $\hat{n}_{i s}=\hat{c}^{\dagger}_{i s}\hat{c}_{i s}$; the {\em
intersite} matrix elements $t_{ij}$ connect nearest neighbor sites and
coincide with the hopping integral $t$ and the {\em on-site} matrix
elements $t_{ii}$ are set equal to the chemical potential $\mu$.

In the functional integral formalism (see, {\it e.g.}, Ref.\
\onlinecite{Vasiliev98}) the generating functional of the Green's 
functions
is [note the absence of the caret on the Grassmann variables below in
contrast to the operators in Eq.\ (\ref{ H})]:
\begin{eqnarray}
\label{ Z0}
&&Z[A,A^+]=\int D\psi \exp\left\{-\int_0^{\beta}d\tau
\left[\sum_i\psi_i^+(\tau)\partial_{\tau}\psi_i(\tau)\right.\right.
\nonumber\\
&&+\left.\left.H(\tau)+\sum_iA_i^+(\tau)\psi_i(\tau)+\sum_i\psi_i^+
(\tau)A_i(\tau)
\right]\right\}
\end{eqnarray}
[cf. Eqs. (29) and (30) of Ref.\ \onlinecite{Georges_ea96}].  Here the
Grassmann variables $A,A^+$ are the spinor source fields which are 
conjugate
to the physical spinor fields corresponding to the electrons
\begin{subequations}
\begin{eqnarray}
\label{ spinors}
&&\psi_i(\tau)
= \left[\begin{array}{c}{c}_{i\uparrow}(\tau)\\{c}_{i\downarrow}(\tau)
\end{array}\right]\\
&&\mbox{and}\nonumber\\
\label{ spinors1}
&&\psi^+_i(\tau)
= [\bar{c}_{i\uparrow}(\tau),\bar{c}_{i\downarrow}(\tau)],
\end{eqnarray}
\end{subequations}
where $\tau$ is the thermodynamic imaginary ``time'', and 
$\beta=1/k_BT$,
where $T$ is the temperature and $k_B$ the Boltzmann constant.  Here 
and below
we shall omit the constants originating from normalizations of
functional integrals, because in the present paper we are interested
only in correlation functions.  The latter are obtained as functional
derivatives of the generating functional (\ref{ Z0}) at zero value of
the functional arguments, divided by $Z[0,0]$, so the normalization is
irrelevant.

If the system is in its normal (i.\ e., non-superconducting) state and,
besides, is paramagnetic, (the case we consider in the present paper),
  the exact one-electron Green's function $G_{ijs}(\tau-\tau^{'})$ as
well as the  self-energy $\Sigma_{ijs}(\tau-\tau^{'})$ are independent
of the spin direction $s=\uparrow,\downarrow$, so we omit this 
subscript for
brevity.  $G$ and $\Sigma$ are
connected via the usual relation which we symbolically write as
\begin{equation}
\label{ G-S}
G=(\partial_{\tau}+\hat{t}-\Sigma)^{-1}.
\end{equation}

The standard diagrammatic analysis (see, e.\ g., Ref.\
\onlinecite{Vasiliev98}) allows one to represent the generating 
functional
of the Green's functions (\ref{ Z0}) as:
\begin{equation}
\label{ Z}
Z[A, A^+]=\exp(A^+GA)R[GA, A^+G].
\end{equation}
To simplify notation, here and below the summations over the site and
spin indices as well as integrations over the imaginary ``time'' are
implicitly assumed in all products of quantities which depend on those
mentioned.  $G$ in Eq.\ (\ref{ Z}) is the exact Green's function
defined in Eq.\ (\ref{ G-S}).  The exponent on the right hand side of
Eq.\ (\ref{ Z}) corresponds to the free propagation of electrons while
the functional $R$ describes their mutual scattering and,
diagrammatically, is represented by the matrix elements of the
$S$-matrix with $GA$ ``tails'' attached.

\section{\em The method} As is well known (see review article
\onlinecite{Georges_ea96} and references
therein), in the infinite dimensional case the DMFT equations are
exact.  In finite dimensions DMFT may be viewed as a single-site
approximation similar to the correlated effective field theory of
ferroelectrics \cite{Lines_ea77} or the coherent potential
approximation in the theory of disordered alloys.\cite{Elliott_ea74}
The latter approximations were found to be quite successful, so the
non-local corrections are expected to be small.  In
Ref.\ \onlinecite{Tokar85} a method of explicit calculation of these
corrections was proposed, based on the observation that in the
self-consistent perturbation theory the Feynman diagrams are expressed
through the exact correlation function (or propagator) $G$.  Thus, if
the propagator is such that its on-site (or site-diagonal) matrix
element is much larger than its off-diagonal elements, then, firstly,
the single-site DMFT-type approximation should be a good one and,
secondly, the dominant correction to this approximation can be
calculated explicitly by taking into account those off-diagonal matrix
elements which are the next largest after the diagonal ones.

It was found that an adequate formalism for the realization of the
above intuitive idea is provided by the functional-differential
quantum-theoretical formalism.\cite{Tokar85}  In this approach the
expression for the generating functional of the $S$-matrix $R$  in Eq.\
(\ref{ Z}) can be cast into the form\cite{Vasiliev98}
\begin{equation}
\label{ R0}
R[\psi, \psi^+]=\exp\left(\frac{\delta}{\delta\psi}{G}
\frac{\delta}{\delta\psi^+}\right)
\exp\left(-\psi^+\Sigma\psi-H_U\right)
\end{equation}
which would coincide with the Hori representation\cite{Hori52} in the
case of $\Sigma=0$.  In the above formula we have used the freedom in
the separation of the total action [the term in the exponent in Eq.\
(\ref{ Z0})] into the ``free'' (bilinear) part and the interaction
part by chosing the (unknown) exact inverse Green's function $G^{-1}$
as the free part. Then according to Eq.\ (\ref{ G-S}) this should be
compensated  by adding the term $\psi^+\Sigma\psi$ to the interaction
part $H_U$. The validity of the above representation can be easily
shown both formally\cite{Tokar97} and through a diagrammatic analysis.

The equation for the unknown exact one-electron Green's function can be
obtained from its definition via the double functional differentiation 
of the generating functional (\ref{ Z0}):
\[G=Z[0,0]^{-1}\delta^2\ln Z[A, A^+]/\delta A\delta 
A^+|_{A, A^+ = 0}.\]
Substituting here the expression (\ref{ Z}) we obtain the following
equation:
\begin{equation}
\label{ the_Eq}
\left.\delta^2\ln R[\psi, \psi^+]/
\delta\psi_i(\tau)\delta\psi^+_j(\tau^{'})\right|_{\psi,\psi^+=0}=0.
\end{equation}

\subsection{The gamma expansion method (GEM) \cite{Tokar85}} The last
equation is exact and completely general (i.\ e., independent of any
approximation scheme).  To solve it in the case of strong correlation
(large $U$) we first note that in Ref.\ \onlinecite{Schindlmayr00} it
was shown that at finite temperature the one electron Green's function
is exponentially damped at large distances.  This confirms the original
suggestion of Ref.\ \onlinecite{Tokar85} about the spatial asymptotic
behavior of the correlation functions as
\begin{equation}
\label{ gamma_def}
G_{ij}\propto\exp(-|i-j|/\xi)=O(\gamma^{|i-j|}),
\end{equation}
where $\gamma=\exp(-1/\xi)$.  Assuming that the correlation length
$\xi$ is small, the Feynman diagrams containing large distance Green's
functions can be neglected and the solution can be approximated by the
local contribution plus a few terms accounting for correlations with
neighboring sites.\cite{Tokar85}From Eq.\ (\ref{ gamma_def}) it
is easy to understand why our approach has DMFT as the zeroth order
approximation.  DMFT is exact in infinite 
dimensions because in dimension $d$ the largest off-diagonal matrix 
elements of the Green function are bounded by a constant of
$O(1/d^{1/2})$ and so vanish at 
$d=\infty$.\cite{Metzner_Vollhardt89,Georges_ea96}  The condition
(\ref{ gamma_def}) with small $\gamma$ presumes that the off-diagonal
elements nearly vanish.   Thus, DMFT should be a good approximation in
such cases in any dimension.

Because the spatial dependence of the self-energy can be expressed 
through
its dependence on the Green's function, we may assume that both $G$ and
$\Sigma$,
considered as matrices in the lattice site indices $i,j$, are diagonally
dominated, so their separation into diagonal and non-diagonal
parts is the separation into the leading contribution and correction
terms:
\begin{equation}
\label{ g+G}
G=gI+\tilde{G},
\end{equation}
where $I=\delta_{ij}$ and $g\equiv G_{ii}$.
Similarly,
\begin{equation}
\label{ s+S}
\Sigma = \sigma I + \tilde{\Sigma}.
\end{equation}
It is easy to check that if in the expressions given in the Appendix
\ref{vertices} one neglects the tilded (i.\ e., intersite) matrix
elements of the Green's functions and of the self-energy operator,
then Eq.\ (\ref{ V}) substituted into Eq.\ (\ref{ the_Eq}) yields the
DMFT equation.\cite{Georges_ea96}

\section{\em Magnetic short-range order}
To illustrate the above general approach we consider the simple problem
of the correlation between electron magnetic moments at nearest neighbor
(NN) sites on the square lattice at half-filling in the insulating
phase.

Following Ref.\ \onlinecite{Okamoto_ea05}, we measure the magnetic 
moment
in Bohr magnetons so, for example, the local moment at site $i$ is 
$m^z_i = n_{i\uparrow}-
n_{i\downarrow}=\hat{c}^{\dagger}_{i}\hat{\tau}^z\hat{c}_{i}$,
where $\hat{\tau}^z$ is the Pauli matrix.  In the paramagnetic phase,
instead of the $\langle m^z_im^z_j\rangle$ correlator used in Ref.\
\onlinecite{Okamoto_ea05} we may consider the equivalent  correlator
\begin{equation}
\label{ mm_mm}
\langle m^-_i(\tau)m^+_j(0)\rangle = 2 \langle m^z_i(\tau) 
m^z_j(0)\rangle
\end{equation}
which is easier to calculate because it contains only the susceptibility
function with opposite spins $\chi_{\uparrow\downarrow}$ while the
former correlator requires also the equal spin susceptibility.  This is 
easily seen from the expression
\begin{equation}
\label{ tau+-}
m_i^-=(m_i^+)^{\dagger}=2\hat{c}^{\dagger}_{\uparrow 
i}\hat{c}_{\downarrow i}
\end{equation}
which follows from the usual definition of the Pauli matrices
$\hat{\tau}^{\pm}=\hat{\tau}^{x}{\pm}i\hat{\tau}^{y}$.

\begin{figure}  %
\begin{center}
\includegraphics[viewport = 0pt 0pt 368pt 102pt, scale = .6]{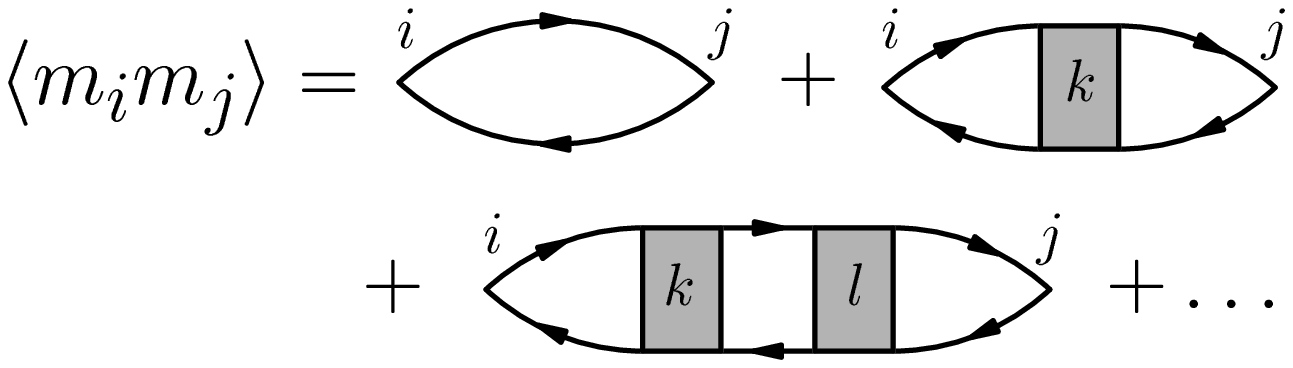}
\end{center}
\caption{\label{fig1}Diagrams contributing into the magnetic short range
order to the lowest $O(\gamma^2)$ order of the small parameter $\gamma$.
$i$ and $j$ are two nearest neighbor sites.
The arrowed lines correspond to the fully renormalized one-electron
Green's function $G$
and the shadowed rectangles correspond to the fully renormalized 
{\em amputated}
single-site vertex $\chi^a_{\uparrow\downarrow}$ (see Appendix
\ref{vertices} for the general definition).}
\end{figure}

We define the magnetic short range order (SRO) parameters $\langle m_i
m_j\rangle$ (we omit the superscripts because they can vary in different
definitions) as the zero-frequency Fourier component of the correlations
functions (\ref{ mm_mm}) with $i\not=j$ which is equivalent to their
average over the interval $0\le\tau\le\beta$.  Because the statistical
average of the product of the operators (\ref{ tau+-}) contains at least
two Green's functions (see Fig.\ \ref{fig1}), the lowest order 
contribution
to the SRO parameter is $O(\gamma^2)$ [see Eq.\ (\ref{ gamma_def})].

A formal expression for $\langle m_i m_j\rangle$ can be obtained from
Eq.\ (\ref{ Z}) by taking the appropriate fourth order functional
derivative with respect to the source variables $A$ and $A^+$.  The
derivative of $R[GA,A^+G]$ should further be expanded
to second order in $\gamma$.  The diagrams containing the
contributions up to $O(\gamma^2)$ are shown in Fig.\ \ref{fig1}.  The
first diagram comes from the first factor in Eq.\ (\ref{ Z})
corresponding to the disconnected contributions.  The second diagram
comes from the zeroth order (DMFT) term of the expansion of $R$ and the
last diagram corresponds to the $O(\gamma^2)$ term in the expansion of
$R$ (the two four-point single-site vertices connected by two intersite
Green's function lines).  Here it is important to note
that the diagrams of Fig.\ \ref{fig1} should not be confused with the
conventional decomposition of two-particle Green's functions into 
ladder-type
series with the shaded rectangles being the irreducible four-point
vertices as is the case with the diagrams shown in Fig.\ 8(b) of Ref.\
\onlinecite{Georges_ea96}.  Despite of their formal similarity, the 
meaning of
the diagrams in the two cases is very different.  While in the 
above-mentioned
Fig.\ 8(b) the shaded rectangles in the case of physical
dimensionalities would correspond to irreducible vertex
functions with their spatial dependence fully taken into account, the
shaded elements in our Fig.\ \ref{fig1} are {\em full} i.\ e., in 
general
{\em reducible} two-particle Green's functions but with their spatial
dependence being only local, i.\ e., restricted to one site ($k$ or
$l$) even in physically relevant finite-dimensional case.  The 
ladder-like structure
of the diagrams shown in Fig.\
\ref{fig1} is purely accidental and will not persist in higher orders.
\subsection{The Hubbard III approximation \cite{Hubbard64}}
\begin{figure}  %
\begin{center}
\includegraphics[viewport = 0pt 0pt 268pt 48pt, scale = .8]{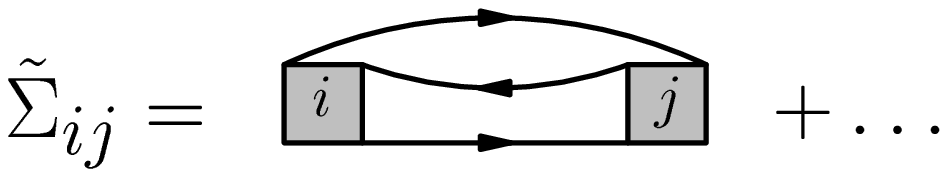}
\end{center}
\caption{\label{fig2}The leading nonlocal correction to the DMFT
self-energy. The meaning of the elements of the diagram is the same as
in the previous figure.}
\end{figure}

To explicitly calculate the above diagrams we need expressions for
their elements: the Green's functions and the four-point vertices.  As
can be seen from Fig.\ \ref{fig2}, the lowest order non-local correction
to the DMFT self-energy $\sigma$ is of order $O(\gamma^3)$, so in 
calculations up
to $O(\gamma^2)$ the exact self-energy $\Sigma$ may be approximated by 
$\sigma$.

For the Hubbard model at half-filling in its
insulating phase the Fermi liquid
properties are not important and the simple Hubbard III approximation
can be used.\cite{Hubbard64} Its detailed discussion, the connection 
with
disordered alloys and the coherent potential approximation (CPA) as well
as further references may be found in Refs.\ \onlinecite{Elliott_ea74}
and \onlinecite{Georges_ea96}.  Here we only mention the observation
made in the latter reference (see sect. VII.C.2) that the Hubbard III
approximation can be obtained as a simple generalization of the atomic
case as follows.
Let us for generality consider the exact atomic Green function at
an arbitrary occupancy $n $\cite{KajueterKotliar96}
\begin{equation}
\label{ g_at}
g_{at}(i\omega)
=\frac{1-n/2}{i\omega+\mu}+\frac{n/2}{i\omega+\mu-U}
\equiv D_{at}[i\omega+\mu-\sigma_{at}(i\omega)].
\end{equation}
If in this equation we replace the inverse unperturbed atomic
Green's function  by
its DMFT counterpart as
\begin{eqnarray*}
g_{0at}^{-1}(i\omega)&&\equiv 
g_{at}^{-1}(i\omega)+\sigma_{at}(i\omega)\\
&&=i\omega+\mu\rightarrow g_{0}^{-1}(i\omega),
\end{eqnarray*}
where, as usual,\cite{Georges_ea96}
\begin{equation}
\label{ g_00}
g_0^{-1}(i\omega)=g^{-1}(i\omega)+\sigma(i\omega),
\end{equation}
the Hubbard III approximation
\begin{equation}
\label{ g_cpa0}
g_s(i\omega) =\frac{1-n/2}{g_0^{-1}(i\omega)}
+\frac{n/2}{g_0^{-1}(i\omega)-U}
\end{equation}
is obtained.

Because the Mott transition takes place only at half-filling
corresponding to $n=1$, the
insulating Hubbard III solution is valid only in this case.
Furthermore, because at half filling the system is explicitly
particle-hole symmetric, it is convenient to switch to an explicitly
particle-hole symmetric formalism.  This is achieved through the
replacement (see Ref.\ \onlinecite{Georges_ea96} Ch.\ VII B)
\[g_0^{-1}\to g_0^{-1}+U/2.\]
In this case Eq.\ (\ref{ g_cpa0}) takes a symmetric form
which we will use in the rest of the paper
\begin{eqnarray}
\label{ g_cpa}
g_s(i\omega) &=&\frac{1/2}{g_0^{-1}(i\omega)+U/2}
+\frac{1/2}{g_0^{-1}(i\omega)-U/2}\nonumber\\
&\equiv& (1/2)g_{s,s}(i\omega)+(1/2)g_{s,-s}(i\omega).
\end{eqnarray}

Here, in the second line, we have introduced the partial single-site
Green's functions corresponding to the propagation of electrons with
their spins parallel or opposite to the dominant spin on the site
under consideration (see Ref.\ \onlinecite{Elliott_ea74}).  These
Green's functions will be necessary in the next subsection.

Eq.\ (\ref{ g_00}) changes to
\begin{equation}
\label{ g_0}
g_0^{-1}(i\omega)=g^{-1}(i\omega)-U/2+\sigma(i\omega).
\end{equation}
Eqs.\ (\ref{ g_cpa}) and (\ref{ g_0}) constitute a set to find
$g(i\omega)$ and $\sigma(i\omega)$, provided the bare single site
Green's function is known.\cite{Georges_ea96}  The latter is given by
the site-diagonal matrix elements of the lattice Green's function and
is usually calculated as a sum over quasimomenta, which co\"{i}ncides
with the Watson integral in the continuum limit\cite{Tokar85}:
\begin{equation}
\label{ D}
D(i\omega) = \frac{1}{N}\sum_{\bf k}\frac{1}
{i\omega-\epsilon({\bf k})},
\end{equation}
where $\epsilon({\bf k})$ is the lattice Fourier transform of the
matrix $-{t}_{ij}$ with $\mu=0$.  With the use of this function
the DMFT Green's function is obtained as $g(i\omega) = D[i\omega
+\mu-\sigma(i\omega)]$.\cite{Georges_ea96}

Similarly, the matrix element $G_1$ of the one-electron Green's 
function connecting
NN sites can be calculated as $G_1=D_1[i\omega+\mu-\sigma(i\omega)]$,
where, for isotropic matrices ${t}_{ij}$ connecting only NN sites:
\begin{equation}
\label{ F1}
D_1(i\omega) = [1-i\omega D(i\omega)]/qt,
\end{equation}
and $q$ is the coordination number ($q=4$ for the square lattice).

We found it convenient to
chose among various forms of the Hubbard III (or CPA)
equations\cite{Elliott_ea74} the following one:
\begin{equation}
\label{ CPA}
\sigma(i\omega)=Un/2+(U/2)^2g_0(i\omega).
\end{equation}
At weak coupling, when the first (Hartree) term in Eq.\ (\ref{ CPA}) 
dominates,
the DMFT approaches the perturbative solution.

\begin{figure}  %
\begin{center}
\includegraphics[viewport = 0pt 0pt 393pt 242pt, scale = .63]{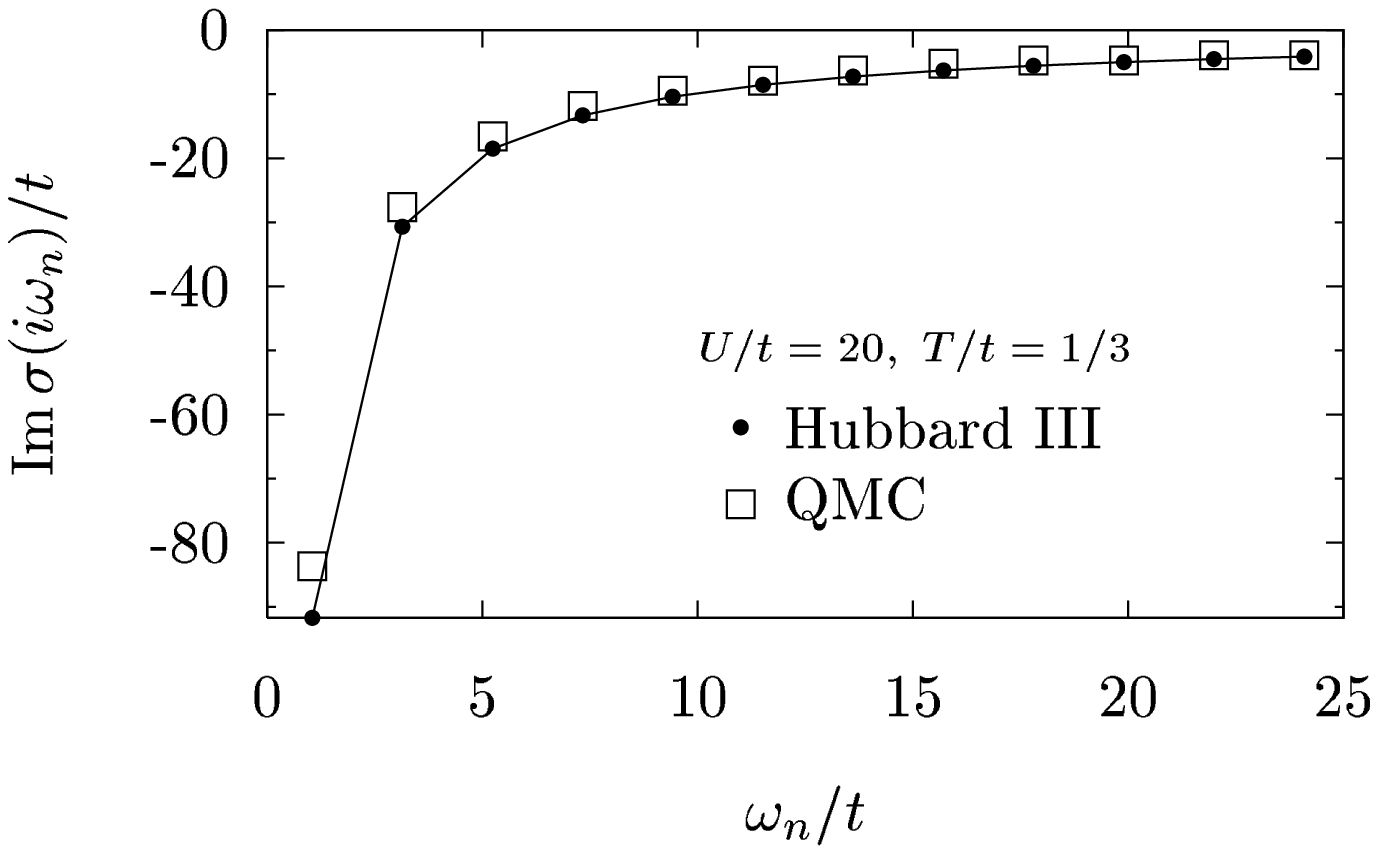}
\end{center}
\caption{\label{fig3}The DMFT self-energy calculated in the Hubbard III
approximation (present work) compared to the quantum Monte Carlo (QMC) 
simulations of Ref.\ \onlinecite{Okamoto_ea05} 
[$\omega_n=\pi k_BT(2n+1)$ is the Matsubara frequency].}
\end{figure}

The solution for $n=1$ was obtained by iterating Eq.\ (\ref{ CPA}) 
together
with Eq.\ (\ref{ g_0}),
and  is shown in Fig.\ \ref{fig3} for a particular set of parameters.  
As we see, the imaginary part of $i\omega_n-\sigma(i\omega_n)$ is quite large,
so the damping with distance of $G_{ij}$ in Eq.\ (\ref{ gamma_def}) is
strong \cite{Schindlmayr00}, hence $\gamma$ is really a small quantity.
For example, with the parameters given in Fig.\ \ref{fig3}
\begin{equation}
\label{ Gmax}
\max|G_1(i\omega_n)|\approx2.5\cdot10^{-3}t^{-1},
\end{equation}
the correction terms being at least of the second order in ${G}_1$.  
This
means that in the case under consideration DMFT is a very good 
approximation.

In the next section we will see that in the case of strong correlation
near half filling there exists a mechanism of enhancement of the loop
diagrams at low temperatures.  At such temperatures the $O(\gamma^3)$
non-local corrections corresponding to the diagram in Fig.\ \ref{fig2}
need to be taken into account.  We will return to this question after
explaining the mechanism of the enhancement.

\subsection{The Curie behavior of the magnetic SRO at low temperatures}

From Eq.\ (\ref{ Gmax}) it follows that deviations from DMFT are 
small unless there exists some mechanism of enhancement of the correction
terms.  In this subsection we are going to show that in the half filled
strongly correlated Hubbard model such a mechanism does exist and is
connected with a peculiar time dependence of the
four-point vertex functions entering the diagrams in Fig.\ \ref{fig1}.
The term exhibiting this dependence is the first one in the expression
for the atomic vertex function Eq.\ (\ref{ chi}) and, at low temperature
\begin{eqnarray}
\label{ v4}
&&\chi^c_{\uparrow\downarrow}(i\omega_1, i\omega_2, i\omega_3,
i\omega_4)\approx\nonumber\\
&&\approx \frac{1}{2}\delta_{\omega_1\omega_4}\delta_{\omega_2\omega_3}
\Delta g(i\omega_3)\Delta g(i\omega_4),
\end{eqnarray}
where the superscript `$c$' stands for `connected', and
\begin{equation}
\label{ del_g}
\Delta g(i\omega)=  
g_{s,s}(i\omega)-g_{s,-s}(i\omega)=-Ug_0(i\omega)g(i\omega).
\end{equation}
All other contributions to $\chi^c_{\uparrow\downarrow}$ do not lead to
an enhancement and can therefore safely be neglected.
By analogy with Eqs.\ (\ref{ g_at})---(\ref{ g_cpa}) above
we may generalize the expression for the vertex
function from the atomic to the band case as in the derivation of the
Hubbard III approximation in the previous subsection, i.\ e., by
replacing the atomic versions of $g_{s,\pm s}$ by their band
counterparts.  This can be justified more rigorously by a calculation
of the vertex function in the Hubbard III
approximation.\cite{1989preprint}

As was pointed out in Ref.\ \onlinecite{Sarker88}, the singular
character of the four-point vertex function is most transparent when 
written
in the form of its contribution to the effective interaction functional
(\ref{ Vexp}):
\begin{equation}
\label{ sarker}
\frac{1}{2}\sum_{\omega}c^{+}_{\uparrow}(\omega)\phi(i\omega)
c_{\downarrow}(\omega)
\sum_{\omega^{'}}c^{+}_{\downarrow}(\omega^{'})\phi(i\omega^{'})
c_{\uparrow}(\omega^{'}),
\end{equation}
where, according to Eqs.\ (\ref{ v4}), (\ref{ del_g}), and (\ref{ vn})
\begin{equation}
\label{ phi}
\phi(i\omega)=Ug_0(i\omega)/g(i\omega)
\end{equation}
[$g^{-1}$ appears due to the four amputations in Eq.\ (\ref{ vn})].
It is easy to see that the above vertex function is a product of two
independent bilinear terms with the symmetry of the $S^{-}$ and $S^{+}$
spin operators.  When transformed to the (imaginary) time the 
interaction
(\ref{ sarker}) takes the form
\begin{equation}
\label{ S-S+}
\frac{1}{2}\int d\tau  S^-(\tau)\int d\tau^{'}S^+(\tau^{'}),
\end{equation}
where the spin variables are the Fourier transforms of the two factors
in Eq.\ (\ref{ sarker}).  Obviously, the above interaction can be {\em
exactly} decoupled with the use of the Hubbard-Stratonovich
transformation with a time-independent ordinary (i.\ e.,
non-functional) variable.  This observation provides a rigorous
justification of the theories based on the static
approximation.\cite{Hubbard79,Hasegawa80,Moriya85}  We stress that,
while in the theories of Refs.\
\onlinecite{Hubbard79,Hasegawa80,Moriya85}  the static approximation
was introduced through an {\em ad hoc} modification of the initial
Hamiltonian, in the present approach it was obtained as an approximate
(Hubbard III) solution to the Hubbard model, based on
the observation made in Ref.\ \onlinecite{Georges_ea96} that,
in the case of strong coupling and half filling, it accurately 
reproduces
the results obtained in a full-fledged DMFT treatment. In particular,
our ``by analogy'' derivation of the vertex function can be confirmed
by the observation that the on-site spin-spin correlation function
``develops long-term memory signalling the formation of a local moment''
made on the basis of numerical simulations [see Ch.\ VII.G.1 of
Ref.\ \onlinecite{Georges_ea96} and Eq.\ (\ref{ B4}) in Appendix
\ref{atomic_v} of the present paper].

Thus, because the static field has a zero-frequency Fourier transform
we see that at strong coupling the half-filled Hubbard model develops a
zero-energy scale.  Below we will show that because of this the
susceptibility may develop a Curie-like behavior on an energy scale
much smaller than the electron bandwidth.  The temperature dependence
$\sim 1/T$ will enhance the contribution of the corresponding diagram
at low temperature,---thus providing a mechanism for the enhancement of
the small $O(\gamma^2)$ contributions mentioned above.

Formally this can be shown as follows.  First, to the order 
$O(\gamma^2)$
we may neglect the difference between $V$ and $\bar{V}$ in Eq.\
(\ref{ R}) because according to Eq.\ (\ref{ tildeV}) their difference
depends on the $O(\gamma^3)$ quantity $\tilde{\Sigma}$.  Thus, the
$O(\gamma^2)$ contribution to $R$ in the spin fluctuation channel
is obtained from Eq.\ (\ref{ R})
as the $O(\tilde{G}^2)$ term in the expansion of the first
exponential.  The contribution (\ref{ sarker}) to $V$ will then lead to 
the the following form for the dominant term in $R$
\begin{equation}
\label{ deltaR}
\delta R =
\frac{1}{2}\alpha(T)\sum_{j\omega}c^{+}_{j\uparrow}(\omega)\phi(i\omega)
c_{j\downarrow}(\omega)
\sum_{k\omega^{'}}c^{+}_{k\downarrow}(\omega^{'})\phi(i\omega^{'})
c_{k\uparrow}(\omega^{'}),
\end{equation}
where $j$ and $k$ denote nearest neighbor sites and
\begin{equation}
\label{ alpha}
\alpha(T) = -U^2\sum_{\omega}[G_1(i\omega)g_0(i\omega)/g(i\omega)]^2.
\end{equation}
Here the minus sign appears because of the closed Fermion line made by 
the two nearest neighbor Green's functions $G_1$.  The remaining factors 
come from the
product of two effective interactions $\phi(i\omega)$.  We note that 
the summation over
the Matsubara frequencies in Eq.\ (\ref{ alpha}) is not supplemented by
a $\beta^{-1}$ factor (this will be explained in more detail below).
Thus, because the $i\omega$-dependent summand consists of smooth
functions and the number of the Matsubara frequencies $\omega_n$ scales
as the inverse temperature, as the temperature is lowered the sum will
diverge as $1/T$.  The importance of this observation lies in the fact 
that this
term leads to the magnetic susceptibility satisfying the Curie law
which we are going to show with explicit calculations below.

Among the diagrams of Fig.\ \ref{fig1} it is the third one which
exhibits the Curie behavior.   This can be seen as follows.  Because
our magnetization operators Eq.\ (\ref{ tau+-}) are dimensionless,
their correlator (\ref{ mm_mm}) or (\ref{ main}) also has no dimension.
The elements of
the diagrams in Fig.\ \ref{fig1}, on the other hand, when Fourier
transformed with respect to the imaginary time variables have the 
following
dimensionalities in units of energy:
\begin{equation}
\label{ dim}
[G] = -1\qquad\mbox{and}\qquad [\chi^a_{\uparrow\downarrow}]=
[\chi^c_{\uparrow\downarrow}][G]^{-4}=2
\end{equation}
[see Eq.\ (\ref{ chi})].
Thus, in the first diagram in Fig.\ \ref{fig1} these elements
contribute the factor $GG$ which has dimension -2.  Because the
contribution should be dimensionless, this means that the diagram
additionally contains the compensating factor $(k_BT)^2$ with the
dimension 2.  According to the diagrammatic rules corresponding to the
vertex function (\ref{ sarker}), in the particle-hole channel
the number of summation over the internal variables of a diagram is 
equal
to the number of loops in this diagram.  The summation over the 
Matsubara
frequencies scales as $\sim 1/T$ for each loop as explained above.
Thus, in the case of the first diagram we have one loop summation of
$O(1/T)$ and the factor $O(T^2)$ so the result is bounded as $T\to0$.
The second diagram has the combination $GG\chi^aGG$ of dimension -2,
hence the same $O(T^2)$ additional factor but this time two loop
summations bringing the $O(1/T^2)$ behavior making the diagram to be
of order unity.  Finally, a similar analysis in the case of the last
diagram in Fig.\ \ref{fig1} shows that it behaves as $O(1/T)$ because
of its three summations over the loop variables.  It is this latter
diagram which are going to compute explicitly below while
neglecting the other two because of their smallness.

The contribution corresponding to this diagram is
\begin{eqnarray}
\label{ main}
&&\langle m^-_im^+_j\rangle \equiv \beta^{-1}\int_0^{\beta}d\tau\langle
m^-_i(\tau)m^+_j(0)\rangle\nonumber\\
&&=m_i m_j\alpha(T),
\end{eqnarray}
where the on-site magnetizations are
\begin{equation}
\label{ m_ij}
m_{i(j)}=-\beta^{-1}\sum_{\omega}\Delta
g(i\omega)=U\beta^{-1}\sum_{\omega}g(i\omega)g_0(i\omega).
\end{equation}
To understand the origin of the contributions to Eq.\ (\ref{ main}) we
first note that the two lines connecting sites $k$ and $l$ on the
diagram come from the intersite correction to the
four-point vertex in Eq.\ (\ref{ R}) where to the order $O(\gamma^2)$ 
we may restrict $\tilde{G}$ to the NN sites.  This explains the
appearance of $G_1^2$ in Eq.\ (\ref{ alpha}).  Thus, the smallness of
the diagram in Fig.\ \ref{fig2} is already exhausted by this
contribution, so the leftmost and the rightmost propagators may
be restricted to on-site Green's
functions $g$.  In other words, to the order $O(\gamma^2)$ we have $i=k$
and $j=l$.  The factors $g^2$ at sites $i$ and $j$ multiply the two
external legs of the vertex (\ref{ v4}) which, combined with Eq.\ (\ref{
phi}), leads to expressions of the type of Eq.\ (\ref{ del_g}). The 
summation over
frequencies then yields the two factors of the type of Eq.\ (\ref{ 
m_ij}).  The on-site
magnetizations $m_{i(j)}$ (\ref{ m_ij}) calculated with the parameters 
of Fig.\ \ref{fig3} are practically saturated to their maximum value 1
(in Bohr magnetons).

In the two internal lines the amputation factors are not compensated by
the propagators because they are now intersite ones ($G_1$), so the two
vertices (\ref{ sarker}) introduce the two factors $Ug_0/g$.
The coefficient
$(1/2)\times(1/2)$ coming from the product of two vertices (\ref{ 
sarker}) is
compensated by the coefficient $2\times2$ coming from Eq.\ (\ref{ 
tau+-}).

\begin{figure}  %
\begin{center}
\includegraphics[viewport = 0pt 0pt 500pt 307pt, scale = .48]{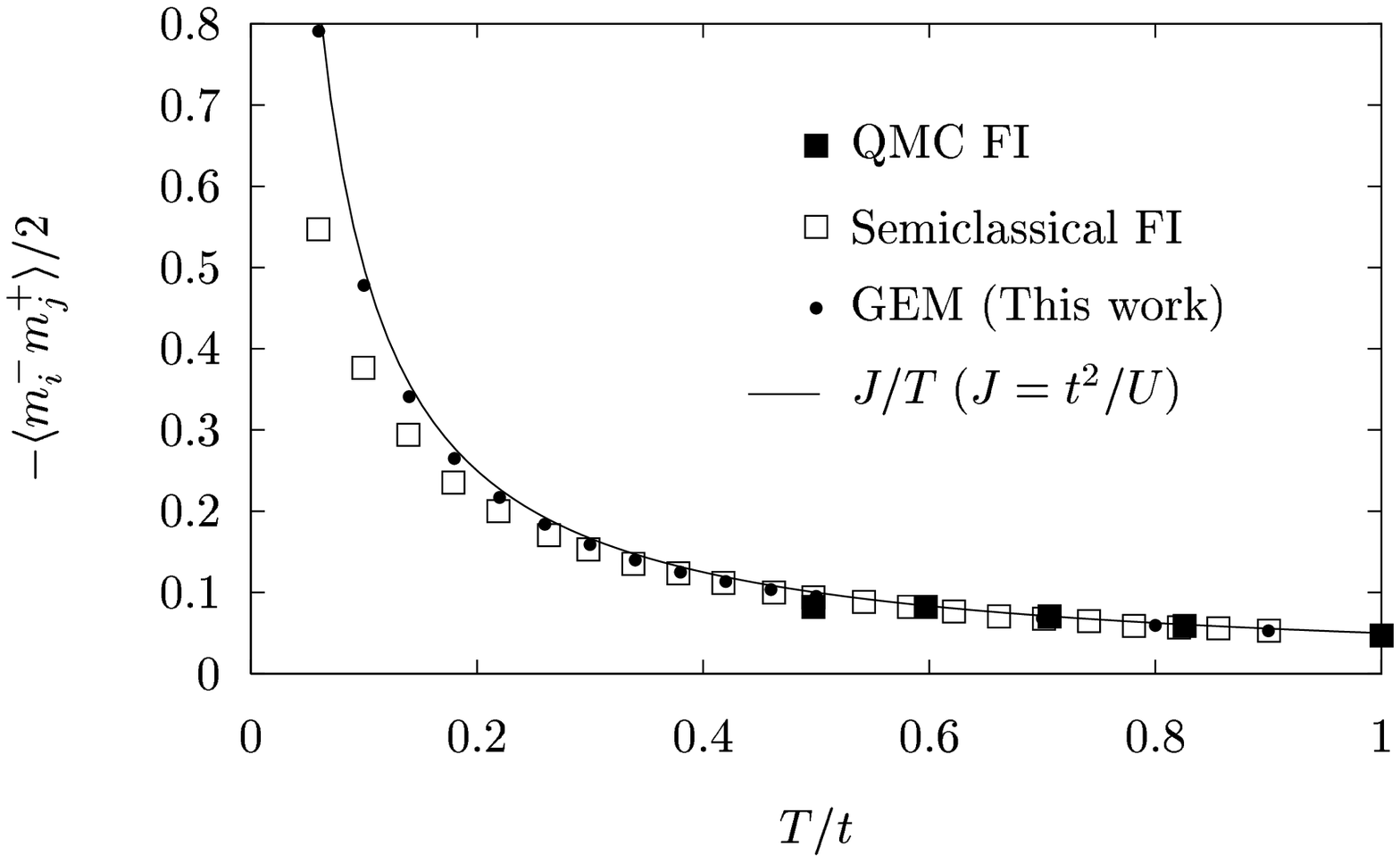}
\end{center}
\caption{\label{fig4}Magnetic short range order parameter for the
NN sites $i$ and $j$ calculated to the lowest nontrivial order of
the gamma expansion method (GEM) in comparison with the cluster 
technique
of Ref.\ \onlinecite{Okamoto_ea03} [called by the authors the fictive
impurity (FI) method] and with the leading term of the high temperature
expansion of the superexchange 
interaction.\cite{Anderson59,Okamoto_ea05}
The square dots correspond to data calculated by two different methods
in Ref.\ \onlinecite{Okamoto_ea05}.}
\end{figure}

The magnetic SRO calculated according to the above formula shows
excellent agreement with the cluster calculations of
Ref.\ \onlinecite{Okamoto_ea05} (see Fig.\ \ref{fig4}).  Because our
technique works directly with the thermodynamic limit, this agreement
confirms the conclusion of the above reference that their fictive
impurity (FI) method is a more reliable cluster approximation than the
dynamical cluster approximation which yields a SRO parameter
approximately two time larger than in Fig.\ \ref{fig4} (see Fig.\ 14 in
Ref.\ \onlinecite{Okamoto_ea05}).

Because of the $O(1/T)$ enhancement, our perturbative technique becomes 
less
reliable at very low temperature, where a resummation technique should 
be used.
For example, by extending the
ladder sum  to infinity with the intermediate
Green's functions restricted to $G_1$ as above, we would have obtained 
the usual mean field value for the Neel temperature $T_N=qJ$.  The value 
of $T_N$ calculated in this way would match that shown in Fig.\ 10 of
Ref.\ \onlinecite{Okamoto_ea05}.  Such an approach, however, is not
in the spirit of the GEM.  In fact, within the GEM the phase
transition temperatures can usually be calculated with much better
accuracy than in the mean field approximation.  For example, according
to Eq.\ (13) of Ref.\ \onlinecite{Tokar85} the value of $qJ/T_N$
in the zeroth order approximation of GEM is given by the Watson
integral for the corresponding lattice.  In a 2-dimensional system this
integral would diverge (see, e.\ g.,
Ref.\ \onlinecite{KatsuraInawashiro71}),---thus giving the correct value
$T_N=0$ in this case. The possibility to develop such an approach for 
phase transitions in the Hubbard model is currently being investigated.

A Curie-like behavior is characteristic for localized spins
but is difficult to understand in the case of itinerant
spins of band electrons.\cite{Fazekas99}  The latter situation may
be described by the Hubbard model in the weak coupling limit $U\to0$.
Because in this case conventional perturbation theory applies, all
calculations can be done in the finite bandwidth case.  But
it is more instructive to see how the perturbative limit is attained
in the formalism of the GEM.  To this end we first note that the
approximation (\ref{ v4}) is not appropriate in the weak coupling case
because it is of second order in $U$. Going back to the defining 
equation
(\ref{ chi}), we find that the first order term has the small-$U$ limit
\begin{equation}
\label{ smallU}
\chi^a_{\uparrow\downarrow}(i\omega_1, i\omega_2, i\omega_3,
i\omega_4)|_{U\to0}
\approx \delta_{\omega_1+\omega_2,\omega_3+\omega_4}k_BTU,
\end{equation}
where we presented the amputated version of the connected correlator
because in this case the atomic correlator is the same as in the band 
case.  As we see, now the vertex functions in the diagrams of Fig.\
\ref{fig1} contain an additional factor $T$, so that the mechanism of
the low-temperature enhancement leading to the Curie law is not
operative any more.  Because of this and because of the smallness of
$U$ it is the first diagram in Fig.\ \ref{fig1} which is the main
contribution with the second diagram giving the first order
correction.  Because the first order Hartree term in the Green
functions is exact and the vertex in Eq.\ (\ref{ smallU}) is also exact,
the spin susceptibility calculated within the GEM will be accurate to
first order in $U$ at weak coupling.
\subsection{Non-local corrections to the self-energy}
The mechanism at the origin of the enhancement in the loop
summations in the half-filled Hubbard model will also be operative in
the diagram for the
lowest-order non-local correction to the self-energy shown in Fig.\
\ref{fig2}.  Indeed, it is easy to see that (i) the contribution to
$\tilde{\Sigma}$ coming from the term (\ref{ deltaR}) can be obtained
by simply contracting $c_{is}$ and $c^+_{js}$ at nearest neighbor
sites and (ii) by noting that the contribution thus obtained should be
multiplied by 3/2 because the term (\ref{ deltaR}) coming from the
product (\ref{ S-S+}) constitutes only 2/3 of the full scalar product
$\vec{S}\cdot\vec{S}=S^-S^++S^zS^z$ constituting the full rotationally
invariant singular interaction \cite{Sarker88} leading to the first 
term in
Eq.\ (\ref{ chi}).  Thus
\begin{equation}
\label{ Sigma}
\Sigma_1(i\omega)=\frac{3}{4}[Ug_0(i\omega)/
g(i\omega)]^2\alpha(T)G_1(i\omega).
\end{equation}
In connection with this expression some comments are in
order.  From Eq.\ (\ref{ alpha}) it follows that the nonlocal 
self-energy in Eq.\ (\ref{ Sigma}) is of fourth order in $U$ .
Such a term was shown to be the lowest order (in $U$)
non-local contribution to the self-energy in the Falicov-Kimball
(F-K) model in Ref.\ \onlinecite{SchillerIngersent95}. Our approximation
(\ref{ v4}) essentially describes this case because it is based on the
local (atomic) approximation to the band electrons---a feature
shared by the $f$-electrons in the F-K model.\cite{FreericksZlatic03}
In the Hubbard model, the last, linear in $U$ term in Eq.\ (\ref{ chi}), 
dominates in the weak coupling limit.  Its insertion into the
diagram of Fig.\ \ref{fig2}  gives the conventional second order 
contribution into $\Sigma$.
\subsection{Herglotz properties of spectral functions}
Our next comment concerns the fundamental requirement that the
imaginary part of the self-energy 
be negative semi-definite in order to guarantee the positivity of the 
density of states 
\begin{equation} 
\label{ rho_k}
\rho(E,{\bf k})\equiv-\frac{1}{\pi}
{\Im \Sigma(E,{\bf k})}{|G(E,{\bf k})|^2}
\end{equation} 
of the quasiparticles with quasimomentum ${\bf k}$ and energy $E$.
The quantities in the right hand side of this equation should be 
obtained by analytic continuation from the discrete Matsubara
frequencies in the upper part of the complex energy plane $z$ 
to the real axis $E$ as
\[z=\left.E+i\epsilon\right|_{\epsilon\to0}.\]
This continuation is in general a non-trivial procedure because the
quantities to be continued are usually known only in the form of the
data of quantum Monte Carlo simulations.  Continuation of such data
to the real axis is an ill-posed problem the discussion of which as
well as the pertinent bibliography may be found in Ref.\
\onlinecite{Georges_ea96}.   We only would like to add that the
continuation of a perturbation expansion (like GEM) poses additional
problems which are explained in Appendix \ref{theorem}.  

Because of the simplicity of the Hubbard III approximation, however,
it is possible to perform the above analytic continuation explicitly.
This will be done below to illustrate the Herglotz properties of the
self-energy and the total density of states with the correction term
given by Eq.\ (\ref{ Sigma}).  To this end we first derive the
expression for the corrected on-site Green's function. 
For a hopping matrix connecting only 
nearest-neighbor sites,  we get from Eqs.\ (\ref{ G-S}) and (\ref{ D})] 
\begin{equation}
\label{ SI}
g(i\omega)\equiv G_{ii}(i\omega)=\frac{1}{1-\Sigma_1(i\omega)/t}
D\left(\frac{i\omega-\sigma(i\omega)+\mu}{1-\Sigma_1(i\omega)/t}\right).
\end{equation}
From this expression it is easy to see that irrespective of the
precise form of the correction term, the total density of states 
(DOS) 
\begin{equation} 
\label{ DOS}
\left.\rho(E)=-\frac{1}{\pi}\Im g(z=E+i\epsilon)\right|_{\epsilon\to0}. 
\end{equation} 
is conserved because the asymptotics of $g(z)$ at large $z$ are the same 
as those of $D(z)$: $\sim1/z$ (see Appendix \ref{theorem} for further 
details).  However, from the point of view of GEM, Eq.\ (\ref{ SI}) is valid 
only to lowest order in
$\Sigma_1$.  So it is more prudent (and also more convenient for
further analysis) to separate different contributions as
\begin{eqnarray} 
\label{ g+dg}
g(i\omega)&\approx& D\left[i\omega-\sigma(i\omega)+\mu\right]
\nonumber\\&+&[\Sigma_1(i\omega)/t]
\left[D(z)+zD^{\prime}(z)\right]_{z=i\omega-\sigma(i\omega)+\mu},
\end{eqnarray} 
where the second line represents the lowest order correction to DMFT
(the first line).

Eqs.  (\ref{ CPA}), (\ref{ Sigma}), (\ref{ SI}), and (\ref{ g+dg})
were continued to the real axis and solved with the model parameters 
used in the calculations above.   The results did not show any violation 
of the Herglotz properties.  
Therefore, to study the anomalies discussed in Appendix \ref{theorem},
a smaller value of $U=3.2t$ was chosen, which corresponds to
80\% of the critical value $U_M\approx4t$ for the Mott
transition in the Hubbard III approximation, and leads
to a more than 30 times larger NN matrix element (\ref{ Gmax})  .

Before proceeding with explicit calculations we would like to remind that
our approach is based on the asymptotic behavior of the imaginary-time
Green's functions at finite temperature.  The spatial attenuation (\ref{
gamma_def}) is defined by the Green's function behavior at the discrete
imaginary energies (the Matsubara frequencies), being more efficient at
larger temperatures.\cite{Schindlmayr00} Therefore, there is no guarantee
that the correction terms will remain small when continued to the real
energy axis.  Explicit calculations below, however, show that the GEM
corrections remain generally small even on the real axis.  

The GEM-corrected self-energy in energy and wave vector space can be written
as [cf.\ Eq.\ (\ref{ s+S})]
\begin{equation} 
\label{ Sigma2}
\Sigma(z,{\bf k})\approx \sigma(z)+2(\cos k_x+\cos
k_y)\alpha\Sigma_1^{\prime}(z),
\end{equation} 
where the prime means that the self-energy (\ref{ Sigma}) has been
divided by the SRO parameter $\alpha(T)$, which is the natural expansion
parameter in the present case.  Since  $\alpha(T)$ is negative, Eq.\
(\ref{ neq}) takes the form
\begin{equation} 
\label{ aneq}
|\alpha| > \left[\max_{\Im\Sigma_1^{\prime}(E, {\bf k})>0}
\frac{2(\cos k_x+\cos k_y)\Sigma_1^{\prime}(E)}{\sigma(E)}\right]^{-1}.
\end{equation} 
An explicit calculation shows that for the case under consideration
the imaginary part of $\Sigma(z,{\bf k})$ is non-negative only 
when $\alpha > -0.45$. We do not consider this
restriction as a severe limitation of the GEM because, in such a situation,
the spins in the system are spatially so strongly correlated, 
that the behavior of the electrons 
changes qualitatively and more sophisticated techniques  should be
developed for calculating the Green's functions 
(see, e.\ g., Ref.\ \onlinecite{Sadovskii_ea05}).

\begin{figure}  %
\begin{center}
\includegraphics[viewport = 0pt 0pt 388pt 411pt, scale = .60]{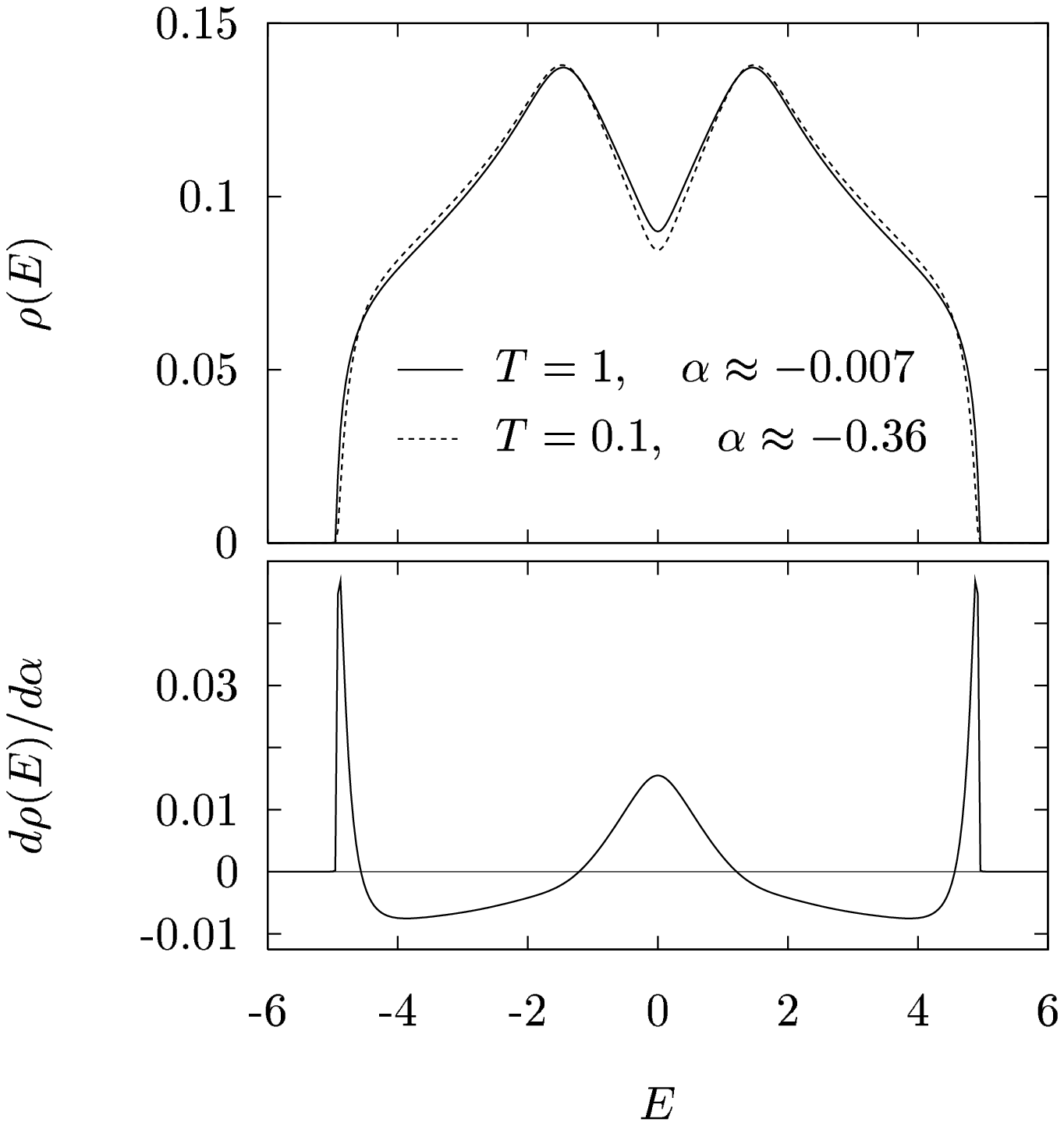}
\end{center}
\caption{\label{fig5}Upper panel: Density of states (DOS) of the 
half-filled Hubbard model on the square lattice with $U=3.2t$ calculated
for two values of the magnetic short range order parameter $\alpha$.
The latter was calculated with the use of Eq.\ (\ref{ alpha}) for the two
temperature shown in the figure.  The DOS was calculated via the
first order expansion Eq.\ (\ref{ g+dg}). Lower panel derivative of the
DOS with respect to the short range order parameter. All quantities 
are measured in energy or inverse energy units, with the
the hopping parameter $t$ set equal to unity.}
\end{figure}

The results of our calculations are shown in Fig.\ \ref{fig5}.  The
Hubbard sub-bands at $U=\pm0.4U_M$ are not yet separated.
From the shape of the derivative of the DOS with
respect to the SRO parameter (lower panel), we expect a
transfer of  spectral weight from the sides of the subbands to their centers
as the temperature is lowered and  $\alpha(T)$ becomes more negative.
The physics of this transfer
is the same as in the Falicov-Kimball (F-K) model of Ref.\
\onlinecite{Hettler_ea00}.  When the local order grows, the atoms
became increasingly surrounded by sites with essentially different
on-site potentials.  In the F-K case these are the atoms filled by 
the $f$-electrons while in the Hubbard case the surrounding atoms
host the electrons of opposite spins. 
This hampers intersite atomic hopping and causes the subband narrowing.
In the F-K model case this can be seen from Fig.\ 7 of Ref.\
\onlinecite{Hettler_ea00}.  The qualitative difference with our Fig.\
\ref{fig5} is only due to (i) the narrow features at the centers of
the subbands because of the atomic-like spectrum of the $f$-electrons
which are absent in the Hubbard model and (ii) stronger temperature
dependence of DOS near the middle of the band because in the 2D F-K model
a charge transfer gap opens in the the spectrum at finite temperature.
This phase transition at finite temperature is possible because the
order parameter of the charge density wave is Ising-like, while in
the 2D Hubbard model the ordering is forbidden due to the continuous
(rotational) symmetry of the spin variables.

Finally, it is worthwhile to note that the enhancement of the intersite
correlations discussed in this work for ``normal'' contractions also takes
place for the anomalous means $\langle c_ic_j\rangle$,\cite{1989preprint}
in which case it leads to a BCS-type superconducting gap equation with
the antiferromagnetic intersite correlations playing the role of the
pairing interaction (see the review paper Ref.\ \onlinecite{Scalapino95}).
This problem will be considered in more detail in a future publication.
\section{\em Conclusion}
As our elementary calculations show, in some cases the GEM allows
for a reliable calculation of non-local corrections to  DMFT. In
the case of the Hubbard model at half filling the corrections are
strongly enhanced by spin fluctuations.  We expect that a similar
(though presumably weaker) enhancement will also take place in the
case of small doping.  Furthermore, from the discussion above, it is
seen that the enhancement occurs in the leading non-local correction
to the four-point vertex (last diagram in Fig.\ \ref{fig1}) and to the
self-energy (particle-hole loops in Fig.\ \ref{fig2}).  In both cases 
the enhancement may be of considerable importance for the spin fluctuation
mechanisms of superconductivity formulated either via the Bethe-Salpeter
equation for the bound states of the pairs or the BCS equation for the
self-energy.\cite{Bulut_ea93,Scalapino95,Maier_ea06} This
  would require more accurate calculation of the four-point vertices
with proper account of the Fermi liquid properties.  This problem
is currently being investigated in the framework of the approaches
of Refs.\ \onlinecite{Hirsch_ea86,Georges_ea96,Rubtsov_ea05}, and
\onlinecite{Troyer_ea05}.
\begin{acknowledgments}
It is a pleasure to thank Matthias Troyer and Emanuel Gull for
enlightening discussions on the continuous time QMC method. One of us
(V. T.) expresses his gratitude to the Center for Theoretical Studies
and the Laboratory for Solid State Physics of the ETH for their kind
hospitality.
\end{acknowledgments}
\appendix
\section{\label{vertices}Renormalized vertices} The relationship 
between the
conventional path integral representation of partition functions and
the Hori functional-differential representation can be formally
established by first using the Gaussian functional integration formula
to decouple the second order functional derivative into a first order
derivative multiplied by a Grassmann field (say, $\eta$) in a way
completely analogous to the Hubbard-Stratonovich transformation and
then applying the formula for the functional shift (Ref.
\onlinecite{Vasiliev98}, Ch.\ 1),
\begin{equation}
\label{ shift}
\exp(\eta\delta/\delta\psi)f[\psi] = f[\psi+\eta]
\end{equation}
valid for any functional $f$ (and similarly for $\eta^{+}, \psi^{+}$).
Next with the sequence of transformations explained
in Refs.\ \onlinecite{Tokar85} and \onlinecite{1989preprint} one
obtains:
\begin{equation}
\label{ R}
R[\psi, \psi^+]=\exp\left(\frac{\delta}{\delta\psi}\tilde{G}
\frac{\delta}{\delta\psi^+}\right)\exp\left(\psi^+\hat{g}^{-1}\psi-
\bar{V}[\psi, \psi^+]\right),
\end{equation}
\begin{equation}
\label{ tildeV}
e^{-\bar{V}[\psi, \psi^+]}=\exp\left(
-\frac{\delta}{\delta\psi}\hat{g}\tilde{\Sigma}\hat{g}\frac{\delta}
{\delta\psi^+} \right)\exp\left(-\sum_iV[\psi_i, \psi^+_i]\right),
\end{equation}
\begin{eqnarray}
\label{ V}
V[\psi_i, \psi^+_i]&=&-\ln\left(\int D\eta D\eta^+\exp\left\{
-\psi^+_i\hat{g}^{-1}\eta-\eta^+\hat{g}^{
-1}\psi_i\right.\right.\nonumber\\
&&+\left.\left.\eta^+(\hat{g}^{-1}+\hat{\sigma})\eta
-H_I[\eta,\eta^+]\right\}\vphantom{\int}\right),
\end{eqnarray}

where
\begin{equation}
\label{ H_I}
H_I[\eta,\eta^+]=\frac{U}{2}\int_0^{\beta}d\tau (\eta^+\cdot\eta)^2,
\end{equation}
and the spinors $\eta$ and $\eta^+$ have the structure
of Eqs.\ (\ref{ spinors}) and (\ref{ spinors1})
\begin{subequations}
\begin{eqnarray}
\label{ spinors2}
&&\eta(\tau)
= \left[\begin{array}{c}{a}_{\uparrow}(\tau)\\{a}_{\downarrow}(\tau)
\end{array}\right]\\
&&\mbox{and}\nonumber\\
\label{ spinors3}
&&\eta^+(\tau)
= [\bar{a}_{\uparrow}(\tau),\bar{a}_{\downarrow}(\tau)].
\end{eqnarray}
\end{subequations}
In Eqs.\ (\ref{ R}) and (\ref{ tildeV}) the summation over the site
indices and the integration over the thermodynamic ``time'' $\tau$ is
implicitly assumed in all products of fields and matrices.  In
Eqs.\ (\ref{ V})--(\ref{ spinors3}) $\eta$ is not supplied with the
site index because we consider the normal, paramagnetic, and
translationally invariant case in which all sites are equivalent.
In this case the matrices $\hat{g}$ and $\hat{\sigma}$ are diagonal
and proportional to the unit matrix because
$g_{\uparrow}=g_{\downarrow}=g$ and similarly for the self-energy.  
The formalism, however, can be easily generalized to a general broken
symmetry case which would essentially amount to introducing 
site-dependent field
$\eta$ and the the Green's function and self-energy matrices of general
form.\cite{Tokar85,1989preprint}

Thus, according to Eq.\ (\ref{ V}) the effective on-site interaction 
vertices are given by
\begin{eqnarray}
\label{ Vexp}
&&V[\psi_i, \psi_i^+]=\sum_{\mbox{even 
$n>2$}}\int_{\tau_1}\int_{\tau_2}\dots
\int_{\tau_{n-1}}\int_{\tau_n}\nonumber\\
&&\psi_i^+(\tau_1)\psi_i^+(\tau_2)\dots
\chi_n^a(\tau_1, 
\tau_2,\dots,\tau_{n-1},\tau_n)\dots\psi_i(\tau_{n-1})\psi_i(\tau_n),
\end{eqnarray}
where the ``amputated" vertex is defined by
\begin{eqnarray}
\label{ vn}
&&\chi_n^a(\tau_1, \tau_2,\dots,\tau_{n-1},\tau_n)
=\int_{\tau^{\prime}_1}\int_{\tau^{\prime}_2}\dots
\int_{\tau^{\prime}_{n-1}}\int_{\tau^{\prime}_n}
\hat{g}^{-1}(\tau_1-\tau^{\prime}_1)\hat{g}^{-1}(\tau_2
-\tau^{\prime}_2)\dots\nonumber\\
&&\langle\eta(\tau^{\prime}_1)\eta({\tau^{\prime}_2})
\dots\eta^+(\tau^{\prime}_{n-1})
\eta^+(\tau^{\prime}_{n})\rangle^c_{S_{eff}}\dots
\hat{g}^{-1}(\tau^{\prime}_{n-1}
-\tau_{n-1})\hat{g}^{-1}(\tau^{\prime}_{n}-\tau_{n}).
\end{eqnarray}
The summation in Eq.\ (\ref{ Vexp}) is restricted to the values of
$n>2$ because according to the zeroth order of GEM (or DMFT---see
below) approximation Eq.\ (\ref{ the_Eq}) with $i=j$ the term $n=2$
turns to zero, so $V[\psi_i, \psi_i^+]$ contains only true
many-electron interactions.  The angular brackets in Eq.\ (\ref{ vn})
denote the statistical averaging in Eq.\ (\ref{ V}) with the effective
action\cite{Georges_ea96}
\begin{equation}
\label{ S_eff}
S_{eff}= \eta^+(\hat{g}^{-1}+\hat{\sigma})\eta -H_I[\eta,\eta^+]
\end{equation}
with the superscript `$c$' meaning that only the connected diagrams
should be kept.  It should be noted that because $\psi_i$ and $\eta$ in
the above equations are of the spinor type, the vertices $\chi_n^a$
are tensor quantities possessing in general $2^{2n}$ components.  In
the symmetric case under consideration many of the components are equal
zero and many among them are mutually equal.  For example, the $n=2$
four-point function is fully defined by two scalar vertices:
$\chi^a_{\uparrow\downarrow}$ and $\chi^a_{\uparrow\uparrow}$ of
which in our calculation of the spin correlation function only the
first one is needed.

Thus, the effective single site vertices in Eq.\ (\ref{ Vexp}) are the
connected [because of the logarithm in Eq.\ (\ref{ V})] amputated
[because of the $\hat{g}^{-1}$ factors in Eq.\ (\ref{ vn})] single-site
correlation functions calculated with the effective action given by
Eq.\ (\ref{ S_eff}).  In Eqs. (\ref{ V}), (\ref{ H_I}), (\ref{ Vexp}),
and (\ref{ vn}) all fields and matrices are assumed to be single-site.
To simplify notation the site dependence has been omitted.  It can be
easily restored by supplying all quantities in these formulas with the
subscript $i$.

There is an infinite number of spatially local interactions in the GEM
but, in contrast to what happens in conventional perturbation theory,
the smallness is provided by the off-diagonal (in the lattice site
indices) elements of the propagator $\tilde{G}$ and of the self-energy
$\tilde{\Sigma}$ [see Eqs.\ (\ref{ R}), (\ref{ tildeV}), and (\ref{ V})
above].  So the order of a correction will be classified by the number
and length of the propagators in the corresponding
diagram.\cite{Tokar85}
For example, if Eq.~(\ref{ the_Eq}) is restricted to a single
site, i. e., if we assume that $i=j$, then Eqs.\ (\ref{ V}), (\ref{
H_I}) with all tilded quantities set to zero together with Eq.\ (\ref{
the_Eq}) will reproduce the DMFT theory of
Ref.\ \onlinecite{Georges_ea96}.  Because $i=j$ means the zeroth order
of the GEM [see Eq.~(\ref{ gamma_def})] the above means that DMFT is
equivalent to the zeroth order of the GEM.
\section{\label{atomic_v}Atomic four-point correlation function}
In this appendix we present the connected part of the atomic spin
susceptibility which was
calculated exactly in the Appendix of Ref.\ \onlinecite{Tokar90} where
this quantity was denoted as $\Gamma_{\sigma,-\sigma}$.  In the present
paper we stick to more common notation of Ref.\ 
\onlinecite{Georges_ea96}.
We consider the limit $U>>k_BT$  and neglect  all terms of the type 
$\exp(-U/2k_BT)$.
Besides, the particle-hole symmetry is introduced by setting the
one-electron energy $\epsilon$ to $-U/2$.
\begin{eqnarray}
\label{ chi}
&&\chi^c_{\uparrow\downarrow}(i\omega_1, i\omega_2, i\omega_3,
i\omega_4)=\nonumber\\
=&&\langle a_{\uparrow}(i\omega_1) a_{\downarrow}(i\omega_2)
\bar{a}_{\uparrow}(i\omega_3)
\bar{a}_{\downarrow}(i\omega_4)\rangle^c\nonumber\\
=&&\prod_{k=1}^4\left(\int_0^{\beta}\frac{d\tau_k}{\sqrt{\beta}}
\right) e^{i\omega_1\tau_1+i\omega_2\tau_2-i\omega_3\tau_3-
i\omega_4\tau_4}\langle a_{\uparrow}(\tau_1)
a_{\downarrow}(\tau_2)\bar{a}_{\uparrow}(\tau_3)
\bar{a}_{\downarrow}(\tau_4)\rangle^c\nonumber\\
=&& 
\frac{1}{2}U^2\left(\delta_{\omega_1\omega_4}\delta_{\omega_2\omega_3}
+\frac{1}{2}\delta_{\omega_1\omega_3}\delta_{\omega_2\omega_4}\right)
f(i\omega_3)f(i\omega_4)\nonumber\\
&&+\frac{1}{2}k_BTUv(\{i\omega_k\})\delta_{\omega_1+\omega_2,\omega_3
+\omega_4},
\end{eqnarray}
where
\[f(i\omega)=\frac{1}{(i\omega)^2-(U/2)^2}\]
and
\begin{eqnarray*}
&&v(\{i\omega_k\})=\left(2\prod_{k=1}^4(i\omega_k+U/2)
-U(i\omega_1+i\omega_2+U)\right.\\
&&\left.\phantom{\prod_{k=1}^4}\times[(i\omega_3+U/2)
(i\omega_4+U/2)+(i\omega_1-U/2)(i\omega_2-U/2)]\right)
\prod_{k=1}^4f(i\omega_k).
\end{eqnarray*}
The atomic approximation may look to be too crude to use in
the finite band width case and in general this is true.  Nevertheless,
it captures in a qualitatively correct way the important particular
cases which we used in our study.  The most important to us
is the first term in Eq.\ (\ref{ chi}) which
we used in our calculations in the main text.  If factor out
the kinematic $\delta$-function responsible for the total energy
conservation as
\begin{equation}
\label{ vertex0}
\frac{U^2}{2}\delta_{\omega_1\omega_4}\delta_{\omega_2\omega_3}
f(i\omega_3)f(i\omega_4)
=\delta_{\omega_1+\omega_2,\omega_3+\omega_4}
\left[\frac{U^2}{2}\delta_{\omega_2-
\omega_3,0}f(i\omega_3)f(i\omega_4)\right]
\end{equation}
one can see that the dynamics of this term is defined by the
zero-energy delta-function singularity in the particle-hole channel
corresponding to the symmetry of the $S^\pm$ spin operators.
A zero-energy excitation in the spin channel is not surprising in
the atomic case where it represents simply the Goldstone boson
due to the rotational symmetry breaking in the ground state in the
half-filled case.  Because of the rotational symmetry
the electron conserves the direction of its spin.

But in the case of itinerant electrons the magnetic moments at
individual sites are not conserved.  Still, the DMFT simulations of
Ref.\ \onlinecite{Georges_ea96} showed that in sufficiently strongly
correlated half-filled Hubbard model the 4-point correlation function
does develop the $\delta$-function singularity in the particle-hole
channel.  This,---in particular,---is reflected in the (imaginary) time
independence of the spin-spin correlation function [see Eq.\ (252) in
Ref.\ \onlinecite{Georges_ea96}].  This can be seen as follows.
According to Eq.\ (\ref{ chi}) [see also Eqs.\ (\ref{ mm_mm}),
(\ref{ tau+-}), and (\ref{ del_g})]
\begin{eqnarray}
\label{ chi_stat}
&&\chi_{loc}^{-+}(\tau)\equiv\langle m_i^-(\tau)m_i^+(0)\rangle
\approx\beta^{-2}\sum_{\omega_1\dots\omega_4}
e^{i(\omega_2-i\omega_3)\tau}\nonumber\\
&&\times 2\delta_{\omega_1\omega_4}\delta_{\omega_2\omega_3}\Delta
g(i\omega_3)\Delta g(i\omega_4)\approx 2,
\end{eqnarray}
where in the last line we used the saturation of the magnetic moments
at strong coupling (see the main text).  Thus, the term under
discussion describes static local magnetic moments because their 
correlation
function is independent of $\tau$.  Because of this, the Fourier
transform has the only component which is different from zero
\begin{equation}
\label{ B4}
\chi_{loc}^{-+}(i\omega=0)=\int_0^{\beta}
d\tau\chi_{loc}^{-+}(\tau)\approx2\beta
\end{equation}
which is approximately twice the result obtain in the QMC simulations
of Ref.\ \onlinecite{Georges_ea96} [see their Fig.\ 44 and Eq.\ (254)]
as it should be in view of Eq.\ (\ref{ mm_mm}).

It should be stressed that the above results are valid only in the {\em
insulating} phase of the model.  The insulator phase appears for $U$
larger than some critical value of $U=U_c$ and only at half-filling.
It is precisely in this case that we apply the approximation (\ref{ 
v4}).

For $U < U_c$ in the band case the $\delta$-function singularity
smears out and transforms itself in one among many others $O(U^2)$
contributions into $v$ in Eq.\ (\ref{ chi}).  In the atomic 
approximation
it will persist at all values of coupling because the atom is always in 
the insulator phase.  But at weak coupling the singularity will be
dumped by the coefficient $U^2$ while the dominant linear in $U$
term given by Eq.\ (\ref{ smallU}) is exact also in the finite band
width case.  Thus, our theory interpolates between two limits where it
agrees with known reliable approaches. 
 \section{\label{theorem}On perturbative expansion of spectral
densities}
Let $A(E)$ be a function corresponding to some spectral density,
which means that the function must be (i)
positive  and (ii) be normalized to some constant value:
\begin{equation} 
\label{ norm}
\int_{-\infty}^{\infty}dE A(E) = C. 
\end{equation} 
The above properties are typical of various spectral functions of the
Green's function theory, such as the densities of states (\ref{
rho_k}) and (\ref{ DOS}) with $C=1$ or
the imaginary part of the electron self-energy multiplied by $-1/\pi$
with $C=U^2n(1-n)$ in the paramagnetic case (see, e.\ g., Appendix A of
Ref.\ \onlinecite{Vilk_ea97}).

Condition (ii) is a consequence of the asymptotic behavior of the
type $C/z$ as $z\to\infty$ of the corresponding complex function of 
which the spectral function is the imaginary part just above
the real axis, multiplied by $-1/\pi$.  

Because the higher order terms of a perturbative expansion usually
contain the products of a larger number of  Green's function than the
lower order terms,  Eq.\ (\ref{ norm}) is usually satisfied in every order
of approximation, provided the order is sufficiently large.  (In the case of the
self-energy, for example, the order should be larger or equal to two.)
In particular, the site-nondiagonal matrix
elements of the Green function which we use in the gamma expansion are 
always quantities of order $o(1/z)$, so the correction terms of any
order cannot modify the $O(1/z)$ asymptotic behavior of a function under
consideration.

Now let us consider some approximation $A^{(0)}\ge0$ to the function
$A(E)$ satisfying Eq.\ (\ref{
norm}).  It is easy to see that any correction term of order $m$,  
$\lambda^m A^{(m)}(E)$ (where $\lambda$ is the expansion parameter) 
should satisfy the restriction
\begin{equation} 
\label{ del_norm}
\int_{-\infty}^{\infty}dE A^{(m)}(E) = 0. 
\end{equation} 
From here it follows that $A^{(m)}(E)$ acquires both positive and
negative values.  Let us for definiteness assume that $\lambda$ is 
positive (the case $\lambda<0$ is treated similarly).  Now if 
\begin{equation} 
\label{ neq}
\lambda^m >
\left[\max_{A^{(m)}<0}\frac{|A^{(m)}(E)|}
{A^{(0)}(E)}\right]^{-1}
\end{equation} 
then, as is easily seen, the corrected 
\begin{equation} 
\label{ Q}
A(E)\approx A^{(0)}(E)+\lambda^mA^{(m)}(E)
=A^{(0)}(E)\left[1+\lambda^m\frac{A^{(m)}(E)}{A^{(0)}(E)}\right]
\end{equation} 
will acquire negative values.

Thus, in perturbative calculations the non-negativity of spectral
functions can be expected only for sufficiently small values of the
expansion parameter.  For larger values special care must be taken
(e.\ g., partial resummation of infinite sequences of the
diagrams) for the calculated spectra to be physically acceptable.  In 
general it should be expected that in a series expansion any exactly 
known property can be guaranteed to be accurate only up to the size 
of the next order correction provided the series converges.  
According to Eq.\ (\ref{ Q}), the condition (\ref{ neq}) 
corresponds to a situation where the correction term becomes larger
than the main term,  which signals a possible divergence of the
series.  In that case a partial sum of the series
cannot be considered as an approximate representation of the function
and usually cannot reproduce its general properties, such as the
positive definiteness.


\end{document}